\documentclass[a4paper,twocolumn,aps,prl]{revtex4-1}
\usepackage{amsmath}
\usepackage{graphicx}
\usepackage{amssymb}

\newcommand{\ket}[1]{\left|#1\right>}
\newcommand{\bra}[1]{\left<#1\right|}

\newcommand{\mb}[1]{\mathbf{#1}}
\newcommand{\Done}{$\text{D}_{1}$ }
\newcommand{\Dtwo}{$\text{D}_{2}$ }
\newcommand{\sigmap}{$\hat\sigma_{+}$ }
\newcommand{\sigmam}{$\hat\sigma_{-}$ }
\newcommand{\pipol}{$\hat\pi$ }

\begin{document}

\title{Two-dimensional imaging of gauge fields in optical lattices}

\author{Jaeyoon Cho}
\author{M. S. Kim}
\affiliation{QOLS, Blackett Laboratory, Imperial College London, London SW7 2BW, UK}

\date{\today}

\begin{abstract}
We propose a scheme to generate an arbitrary Abelian vector potential for atoms trapped in a two-dimensional optical lattice. By making the optical lattice potential dependent on the atomic state, we transform the problem into that of a two-dimensional imaging. It is shown that an arbitrarily fine pattern of the gauge field in the lattice can be realized without need of diffraction-limited imaging.
\end{abstract}

\maketitle

Recently, many-body systems of trapped atoms have offered a new avenue towards understanding strongly correlated matters \cite{greiner02}. In this context, trapped atoms in gauge fields have attracted much attention \cite{abo01,*lin09,jaksch03,*sorensen05,*osterloh05,*williams10,mueller04,*gerbier10,palmer06}. In a two-dimensional (2D) confinement, such systems would exhibit intriguing phenomena, such as the fractional quantum Hall effect and the anyonic statistics, which have opened up a new era of modern condensed matter physics \cite{laughlin83,wen04}. 

The aim of this paper is to propose a concrete framework to realize the following Hamiltonian in an optical lattice system
\begin{equation}
H=-J_{0}\sum_{j,k}\bigl(c_{j+1,k}^{\dagger}c_{j,k}e^{i\theta_{j,k}}+c_{j,k+1}^{\dagger}c_{j,k}+\text{H.c.}\bigr),
\label{eq:peierl}
\end{equation}
which describes a tight-binding particle in the lowest band of a 2D square lattice in a spatially slow-varying magnetic field $\mb{B}(\mb{r})=\mb{\triangledown}\times\mb{A}(\mb{r})$ \cite{luttinger51}. Here, the positions of lattice sites are $\mb{r}_{j,k}=r_{0}(j\hat{x}+k\hat{y})$ with $r_{0}$ being the lattice spacing, $c_{j,k}$ denotes a particle annihilation operator at the site, and $\theta_{j,k}=(2\pi/\Phi_{0})\int_{\mb{r}_{j,k}}^{\mb{r}_{j+1,k}}\mb{A}\cdot d\mb{l}$ with $\Phi_{0}=h/e$ being the magnetic flux quantum. We have chosen a gauge such that the vector potential is written as $\mb{A}(\mb{r})=A(x,y)\hat{x}$. While the lattice structure gives rise to intriguing physics in its own right, such as Hofstadter butterfly \cite{hofstadter76}, this model Hamiltonian also restores the fractional quantum Hall physics in the continuum limit when a strong on-site interaction is considered in addition \cite{palmer06}. Such an optical lattice system would thus provide an ideal testbed to study the physics in the presence of gauge fields, along with established techniques for its microscopic control and measurement \cite{bloch08,bakr10,*sherson10}.

Before proceeding further, we highlight the underlying idea and important features of the present work. Differently from earlier schemes in this context \cite{jaksch03,*sorensen05,*osterloh05,*williams10, mueller04,*gerbier10}, our scheme is devised to make it feasible to realize an arbitrary $\mb A(\mb r)$, which enables the generation of quasiparticles/quasiholes by adiabatically changing $\mb A(\mb r)$ \cite{laughlin83}. This may open up an intriguing possibility of directly and microscopically inspecting the fractional charge and statistics of the quasiparticles, as well as other modes of measurement discussed in earlier literature \cite{jaksch03,*sorensen05,*osterloh05,*williams10}. Furthermore, we achieve this goal along with overcoming the earlier experimental problems, such as the stability issue or complicated experimental setup \cite{jaksch03,*sorensen05,*osterloh05,*williams10,mueller04,*gerbier10}. Unlike most of the earlier schemes based on engineering Raman-induced hopping, we simply make use of the ordinary tunneling in an optical lattice. We employ the state-dependent lattice potential that naturally comes about by adjusting the detuning and polarization of the trapping laser \cite{mandel03,*liu04}. On top of this simple setup, we show that the spatial {\em phase} profile of a {\em static} driving field, which induces on-site Raman transition of atoms with different phases, is mapped into the Abelian gauge field $\mb A(\mb r)$. Our scheme should thus be well within state-of-the-art technology and compatible with currently prevalent setups based on alkali-metal atoms. Moreover, as we use the atomic dark states decoupled from the Raman field, the recoil heating is minimized. While the diffraction limit seems problematic in imaging an arbitrary phase pattern site-by-site on the lattice, we show that a 2D array of only moderately well focused beams, each centered on a lattice site, in fact suffices for any $\mb A(\mb r)$, aside from the available technology of subdiffraction imaging \cite{liu07}.

\begin{figure}
\begin{tabular}{cc}
\includegraphics[width=0.36\columnwidth]{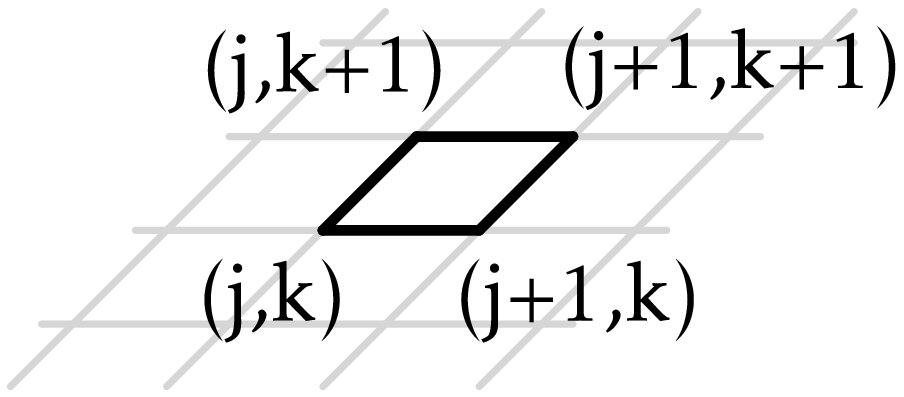} &
\includegraphics[width=0.36\columnwidth]{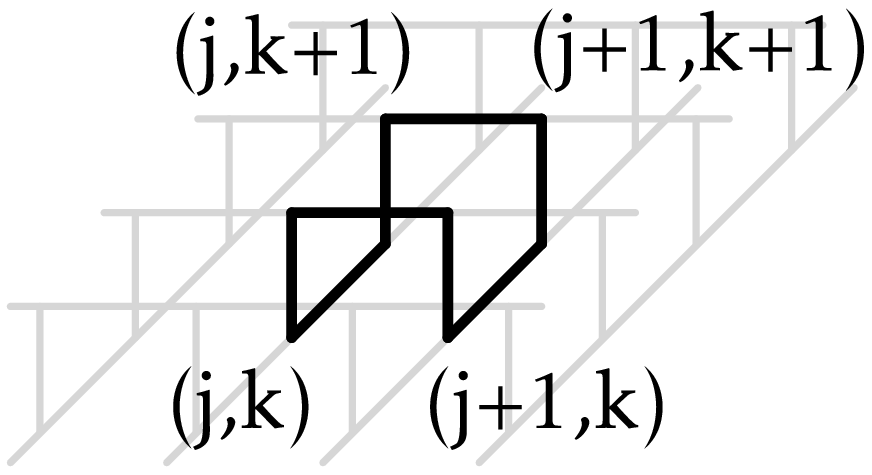} \\
(a) & (b)
\end{tabular}
\caption{The smallest loop in a square lattice (a) and a bilayer lattice we consider (b). In (b), the upper and lower layers trap atoms in $\ket{a}$ and $\ket{b}$, respectively. Both loops enclose the same number of magnetic flux quanta (mod 1) $(\theta_{j,k}-\theta_{j,k+1})/2\pi$.}
\label{fig:bilayer}
\end{figure}

We first present our (abstract) Hubbard model, putting off the detailed explanation of its realization until later. Let us consider a situation where two hyperfine levels $\ket{a}$ and $\ket{b}$ of an atom experience different optical lattice potentials, while the two lattices coincide. In particular, let us assume the atoms in $\ket{a}$ ($\ket{b}$) hop only in the $\hat{x}$ ($\hat{y}$) direction. The Hamiltonian for this state-dependent hopping can be written as
\begin{equation}
H_{l}=-J\sum_{j,k}\bigl(a_{j+1,k}^{\dagger}a_{j,k}+b_{j,k+1}^{\dagger}b_{j,k}+\text{H.c.}\bigr),
\label{eq:setup}
\end{equation}
where $a_{j,k}$ ($b_{j,k}$) denotes the annihilation operator for atoms in $\ket{a}$ ($\ket{b}$) and for simplicity the hopping rate $J$ is assumed to be the same for each direction. On top of this, Raman fields for transition $\ket{a}\leftrightarrow\ket{b}$ are applied with position-dependent phases $\phi_{j,k}$. Note that these phases are determined unambiguously by the difference between the phases of two Raman fields. The Hamiltonian for the Raman transitions is written as
\begin{equation}
H_{r}=\omega\sum_{j,k}\bigl(a_{j,k}^{\dagger}b_{j,k}e^{-i\phi_{j,k}}+\text{H.c.}\bigr),
\label{eq:raman}
\end{equation}
where $\omega$ is the (real-valued) Raman transition rate assumed to be the same for every lattice site. For convenience, let us perform a local gauge transformation $a_{j,k}\rightarrow a_{j,k}e^{-i\phi_{j,k}}$. The single-particle Hamiltonian $H_{s}=H_{l}+H_{r}$ then reads
\begin{equation}
\begin{split}
H_{s}=&-J\sum_{j,k}\bigl(a_{j+1,k}^{\dagger}a_{j,k}e^{i\theta_{j,k}}+b_{j,k+1}^{\dagger}b_{j,k}+\text{H.c.}\bigr)\\
&+\omega\sum_{j,k}\bigl(a_{j,k}^{\dagger}b_{j,k}+\text{H.c.}\bigr),
\end{split}
\label{eq:bilayer}
\end{equation}
where $\theta_{j,k}=\phi_{j+1,k}-\phi_{j,k}$. Note that $\theta_{j,k}$ can be chosen arbitrarily and independently for each pair of sites. We are going to choose $\theta_{j,k}$ as defined in Eq. \eqref{eq:peierl} (conversely, the laser phases determine the vector potential $\mb{A}(\mb{r})$ of the simulated system). Note that the smallest loop of this system as shown in \figurename~\ref{fig:bilayer}(b), which is not 2D, encloses the same amount of magnetic fluxes as that in \figurename~\ref{fig:bilayer}(a).

\begin{figure*}
\includegraphics[width=0.99\textwidth]{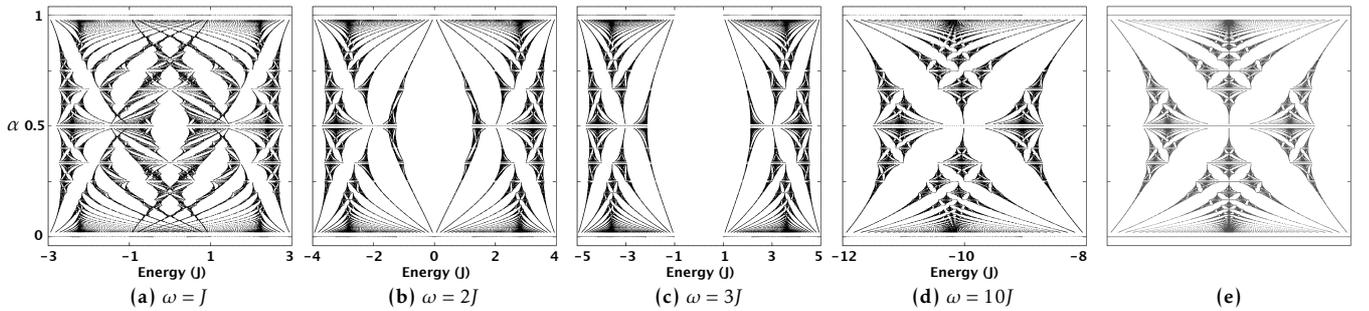}
\caption{The energy spectrum of the system for different parameter choices. In (d), only one of the two bands, with the energy ranging from $-12J$ to $-8J$, is plotted, which resembles the original Hofstadter butterfly shown in (e).}
\label{fig:hofstadter}
\end{figure*}

A further progress is made by performing a transformation such that $c_{j,k}\equiv\frac{1}{\sqrt{2}}(a_{j,k}-b_{j,k})$ and $d_{j,k}\equiv\frac{1}{\sqrt{2}}(a_{j,k}+b_{j,k})$. Eq. \eqref{eq:bilayer} then reads $H_{s}=H_{0}+H_{1}$, where
\begin{align}
H_{0}=&-J_{0}\sum_{j,k}\bigl(c_{j+1,k}^{\dagger}c_{j,k}e^{i\theta_{j,k}}+c_{j,k+1}^{\dagger}c_{j,k}+\text{H.c.}\bigr)\nonumber\\
&-J_{0}\sum_{j,k}\bigl(d_{j+1,k}^{\dagger}d_{j,k}e^{i\theta_{j,k}}+d_{j,k+1}^{\dagger}d_{j,k}+\text{H.c.}\bigr)\nonumber\\
&+\omega\sum_{j,k}\bigl(d_{j,k}^{\dagger}d_{j,k}-c_{j,k}^{\dagger}c_{j,k}\bigr),\\
H_{1}=&-J_{0}\sum_{j,k}\bigl(c_{j+1,k}^{\dagger}d_{j,k}e^{i\theta_{j,k}}+d_{j+1,k}^{\dagger}c_{j,k}e^{i\theta_{j,k}}+\text{H.c.}\bigr)\nonumber\\
&+J_{0}\sum_{j,k}\bigl(c_{j,k+1}^{\dagger}d_{j,k}+d_{j,k+1}^{\dagger}c_{j,k}+\text{H.c.}\bigr),
\end{align}
where $J_{0}\equiv J/2$ is chosen in accordance with Eq. \eqref{eq:peierl}. 
It can be seen that the energy spectrum of $H_{0}$ alone is divided into two bands centered at $\pm\omega$, each corresponds exactly to that of our desired Hamiltonian \eqref{eq:peierl}, whereas $H_{1}$ only contains terms that give rise to transitions between the two bands. These transitions are energetically costly for $2\omega\gg J_{0}$, in which case they are suppressed and hence $H_{1}$ can be treated as a perturbation. This becomes clearer when we consider a uniform magnetic field $\theta_{j,k}=2\pi\alpha k$, where $\alpha$ is the number of magnetic flux quanta passing through the lattice cell. In \figurename~\ref{fig:hofstadter}, we plot the energy spectrum of Hamiltonian \eqref{eq:bilayer}. The horizontal axis represents the energy and the vertical axis  $\alpha$, where rational numbers $\alpha=p/q$ with $p$, $q$ coprime integers and $q<50$ are taken. As the width of each energy band is $4J$, two energy bands are separated when $\omega>2J$. As is expected, for a large $\omega$ (e.g., see \figurename~\ref{fig:hofstadter}(d)), the original shape of the Hofstadter butterfly is restored in each band, albeit slightly deformed due to the perturbation. This deformation gets smaller for larger $\omega$. 

For a system of $N$ atoms, one can choose the lowest energy band centered at $-N\omega$, wherein all the particles are approximately in $c_{j,k}$ modes. It is noteworthy that $c_{j,k}$ particles are in dark states when the corresponding two Raman fields have the same Rabi frequency. As these particles are decoupled from the Raman fields, they are, in principle, not influenced by the spontaneous emission. In order to see that this system can produce the fractional quantum Hall physics, we again employ a uniform magnetic field $\theta_{j,k}=2\pi\alpha k$ and consider the following on-site interaction Hamiltonian on top of the above single-atom Hamiltonian:
\begin{equation}
H_{i}=U\sum_{j,k}\bigl[(a_{j,k}^{\dagger})^{2}a_{j,k}^{2}+(b_{j,k}^{\dagger})^{2}b_{j,k}^{2}+a_{j,k}^{\dagger}a_{j,k}b_{j,k}^{\dagger}b_{j,k}\bigr],
\end{equation}
where for simplicity all the interaction rates are assumed to be the same as $U$. As an example, we have numerically diagonalized the Hamiltonian for two atoms in a $8\times8$ lattice, where a periodic boundary condition with respect to the magnetic translation is taken so as to account for a large lattice system \cite{haldane85a,*haldane85b}. We have taken $U=\omega=10J$ and $\alpha=1/16$, hence the filling factor is $\nu=1/2$. Due to the torus geometry, this system has twofold degenerate ground states. We have obtained the two lowest energy eigenstates $\rho_{\mu}$ ($\mu=1,2$) by tracing out the internal state. The purities of these states are $\text{Tr}(\rho_{1}^{2})=0.994$ and $\text{Tr}(\rho_{2}^{2})=0.993$, which means the internal degree of freedom is almost decoupled from the motional degree of freedom, and the number of particles in $c_{j,k}$ modes is $\text{Tr}(\sum c_{j,k}^{\dagger}c_{j,k}\rho_{\mu})=2.000$. These are in good agreement with the discussion above. We also define a projector to the subspace spanned by the two Laughlin ground states on the torus $P_{L}=\ket{\Psi_{1}^{L}}\bra{\Psi_{1}^{L}}+\ket{\Psi_{2}^{L}}\bra{\Psi_{2}^{L}}$ and calculate the overlap of the above numerical ground states to this subspace, which is found out to be $\text{Tr}(P_{L}\rho_{\mu}P_{L})=0.999$. We also consider second-nearest-neighbor hopping in case the lattice potential in the direction of hopping should be set relatively shallow. As this does not destroy the structure associated with $\alpha$, the ground state is not expected to be altered significantly. Our numerical calculation indicates that for the second-nearest-neighbor hopping rate set to be $J/10$, the overlap of the ground state is still as high as 0.998.

\begin{figure}
\begin{tabular}{cc}
\includegraphics[width=0.38\columnwidth]{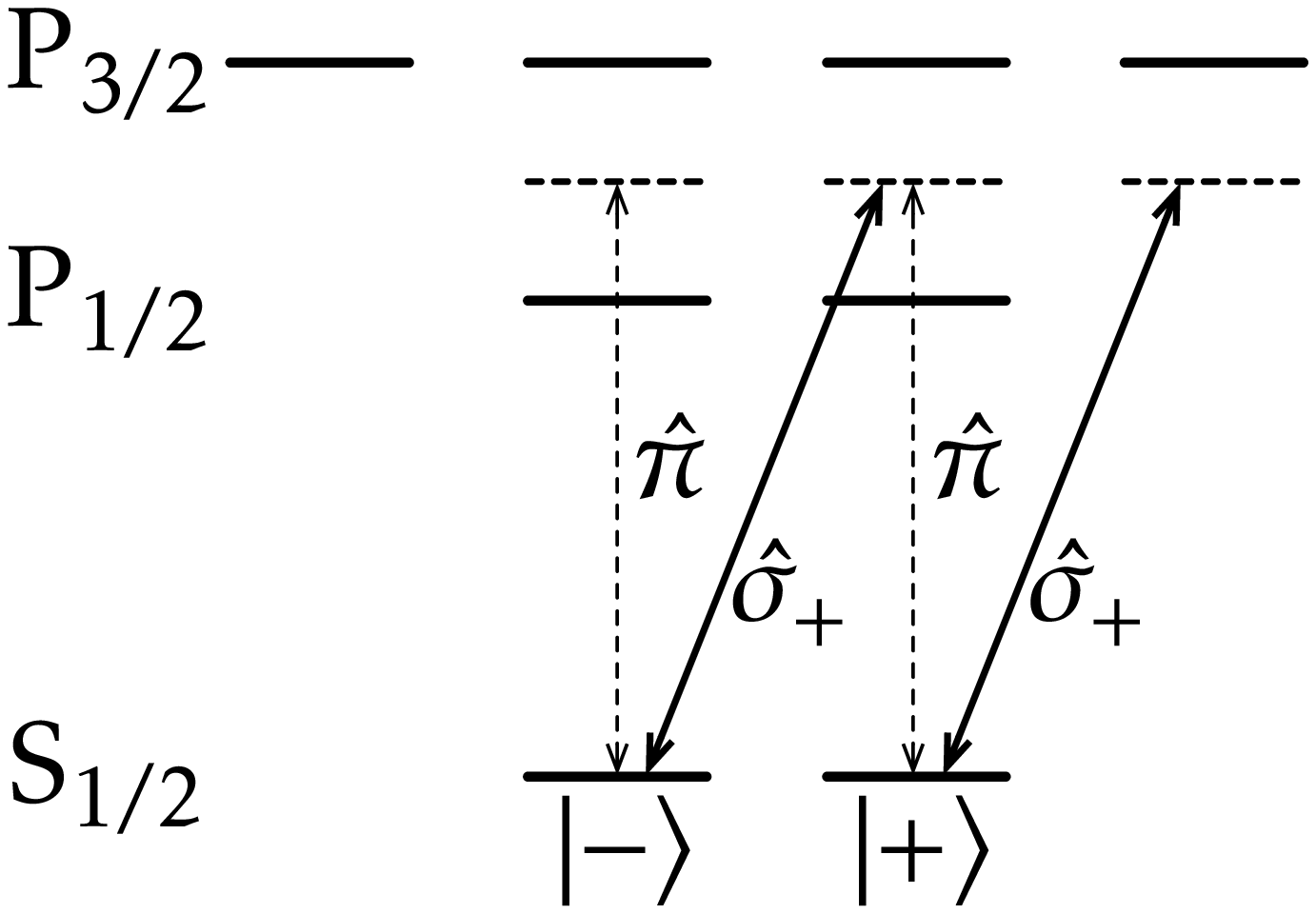} &
\includegraphics[width=0.38\columnwidth]{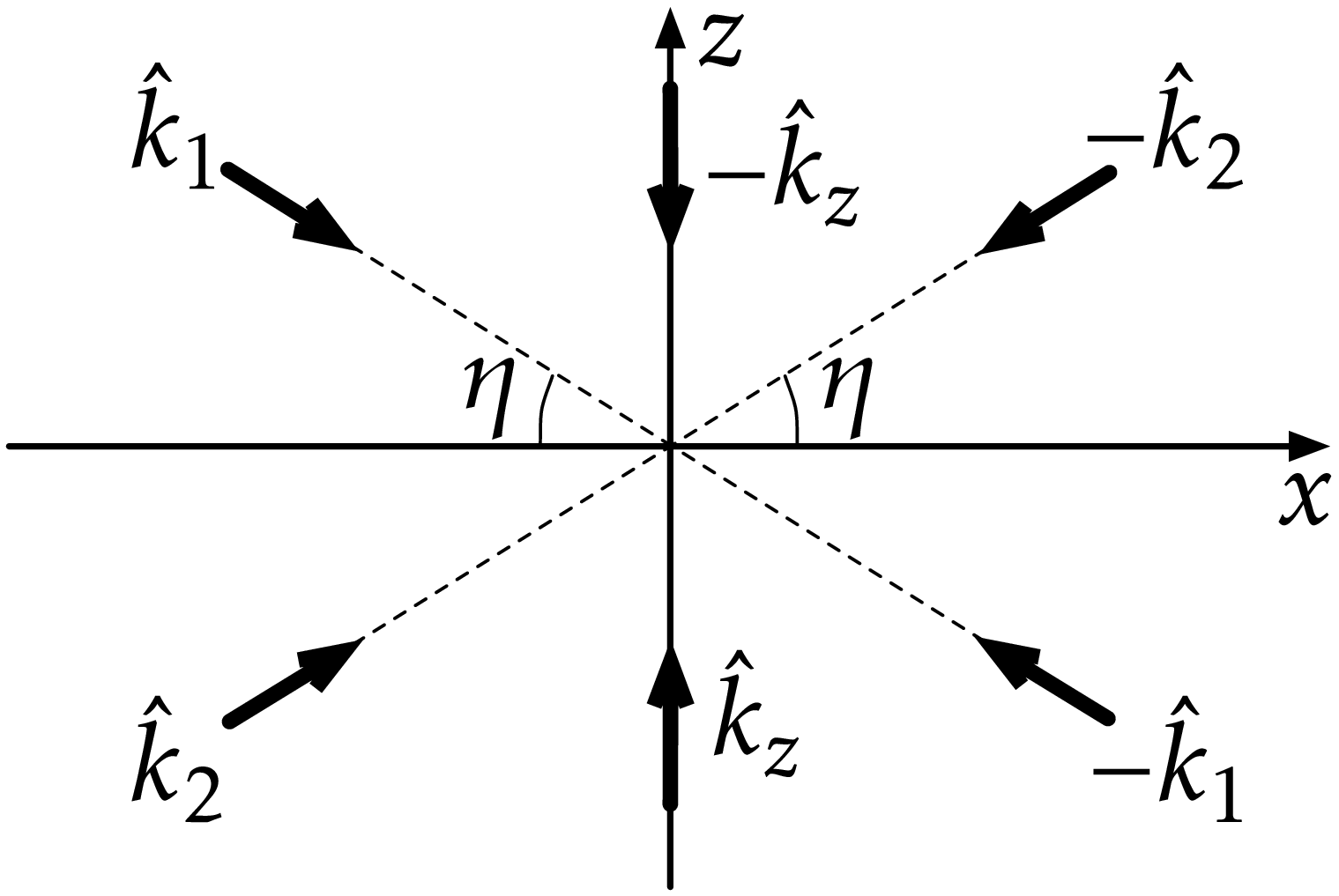} \\
(a) & (b)
\end{tabular}
\caption{(a) The fine structure of alkali-metal atoms, where the transitions by $\hat\sigma_{-}$ polarization, equivalent to those by $\hat\sigma_{+}$ with $m_{J}\rightarrow-m_{J}$, are omitted for brevity, and (b) a laser configuration for the lattice potential in $\hat x$ direction.}
\label{fig:setup}
\end{figure}

The remaining question is how to realize Hamiltonians \eqref{eq:setup} and \eqref{eq:raman}. To be concrete, we introduce below a particular setup based on alkali-metal atoms, generalizing the conventionally used one-dimensional (1D) setups \cite{mandel03,*liu04}. However, the idea would be applicable to different atomic species or laser configurations.

Hamiltonian \eqref{eq:setup} requires the lattice potential to depend on the atomic state. This can be achieved by exploiting the polarization dependence of atomic dipole transitions. Let us first consider the hopping in $\hat x$ direction. The basic idea is as follows \cite{mandel03,*liu04}. Consider the fine structure of alkali-metal atoms and suppose that the frequency of the trapping laser is chosen between those of \Done ($\text{S}_{1/2}\leftrightarrow\text{P}_{1/2}$) and \Dtwo ($\text{S}_{1/2}\leftrightarrow\text{P}_{3/2}$) transitions,  as shown in \figurename~\ref{fig:setup}(a). If the trapping laser is in \sigmap polarization, atoms in $\ket+$ would experience a negative ac Stark shift $V_{+}$ as the laser is red-detuned from the \Dtwo transition. On the other hand, atoms in $\ket-$ would experience both negative and positive ac Stark shifts due to the \Dtwo and \Done transitions, respectively. When these two Stark shifts are summed, the resulting energy shift $V_{-}$ can range from negative to positive values, depending on the laser frequency, while $V_{+}$ is still negative in any case. The ratio $V_{+}/V_{-}$ can thus be adjusted to a great extent. The energy shift for each hyperfine level is now obtained as a linear combination of $V_{\pm}$, determined by the Clebsch-Gordan coefficients. As an example, let us take $\ket{a}\equiv\ket{F=1,m_{F}=+1}$ and $\ket{b}\equiv\ket{F=2,m_{F}=+1}$ among the ground hyperfine levels of ${}^{87}\text{Rb}$. The ratio between the corresponding lattice potentials is then given by $V_{b}(x)/V_{a}(x)=(3V_{+}+V_{-})/(V_{+}+3V_{-})$. For example, suppose the frequency of the laser is chosen such that $V_{+}=-7V_{-}$, leading to $V_{b}(x)/V_{a}(x)=5$ (note that this can be done when the frequency is red-detuned compared to that corresponding to $V_{-}=0$). In 1D cases, the hopping rate is proportional to $E_{r}(V_{0}/E_{r})^{3/4}\exp(-2\sqrt{V_{0}/E_{r}})$, where $E_{r}=\hbar^{2}k^{2}/2m$ is the recoil energy and $V_{0}$ is the potential depth \cite{bloch08}.  If we take moderately $V_{a}(x)=-5E_{r}$ at the minima, the above choice leads to a hopping rate for $\ket{b}$ being 0.013 times that for $\ket{a}$ in $\hat x$ direction, well approximating our model. 

In an analogous fashion, one could use \sigmam polarization in $\hat y$ direction. In a 2D geometry, however, this cannot be done without introducing \pipol polarization when the beam in $\hat x$ direction is chosen to be \sigmap polarized. When the \pipol polarization is involved, as well as the ac Stark shift, a Raman transition can also take place, as can be seen in \figurename~\ref{fig:setup}(a). Although the Raman transition could be suppressed by applying a magnetic field, it would be advantageous to exploit the large fine splitting to obtain a stronger potential, as we discuss below. Let us take the convention that $\hat\sigma_{\pm}=\frac{1}{\sqrt2}(\hat x\pm\hat y)$ and $\hat\pi=\hat z$. In \figurename~\ref{fig:setup}(b), we depict the laser configuration in $\hat x$ direction. The lattice potential in $\hat y$ direction can be created analogously. In $\hat x$ direction, two tilted standing waves with wavevectors $\pm\hat k_{1}=\pm k(\cos\eta\hat x-\sin\eta\hat z)$ and $\pm\hat k_{2}=\pm k(\cos\eta\hat x+\sin\eta\hat z)$ and polarizations $\hat\epsilon_{1}\propto\sqrt{2}\sin\eta\hat\sigma_{+}+\cos\eta\hat z$ and $\hat\epsilon_{2}\propto\sqrt{2}\sin\eta\hat\sigma_{+}-\cos\eta\hat z$, respectively, are applied (note $\hat k_{1}\cdot\hat\epsilon_{1}=\hat k_{2}\cdot\hat\epsilon_{2}=0$). Choosing the phases appropriately, the electric field can be written as $\mb{E}(\mb r)=\hat\sigma_{+} E_{+}(\mb r)+\hat\pi E_{\pi}(\mb r)$ with $E_{+}(\mb{r})\propto\sqrt{2}\sin\eta\cos(kx\cos\eta)\cos(kz\sin\eta)$ and $E_{\pi}(\mb{r})\propto\cos\eta\sin(kx\cos\eta)\sin(kz\sin\eta)$. Note that $E_{+}(\mb r)$ and $E_{\pi}(\mb r)$ are $\pi/2$ out of phase in both directions. The minima of the potential for $\hat\sigma_{+}$ ($\propto-|E_{+}(\mb r)|^{2}$) correspond to the maxima of the potential for \pipol ($\propto-|E_{\pi}(\mb r)|^{2}$), and vice versa. If we choose a 2D plane formed, e.g., at $z=0$ with a strong confining potential in $\hat z$ direction, the influence of the \pipol polarization to the lattice potential is negligible. Furthermore, denoting by $W_{\pm}(\mb r)$ the (real and symmetric) Wannier functions for state $\ket\pm$, the Raman transition rate $\propto\bigl|\int d\mb r E_{+}(\mb r)E_{\pi}(\mb r)W_{+}(\mb r-\mb r_{j,k})W_{-}(\mb r-\mb r_{j,k})\bigr|$ always vanishes for every site $\mb r_{j,k}$ because the electric field part is an odd function around $\mb r_{j,k}$. Note that the lattice spacing is $1/\cos\eta$ times larger than usual $\lambda/2$ because the lasers are tilted. For the same $V_{0}$, this decreases the hopping rate, hence the energy scale of the system. Whereas this can be mitigated by decreasing $\eta$, smaller $\eta$ results in a shallower potential. This trade-off relation would be important in choosing $\eta$ in experiments.

Finally, we discuss the realization of Hamiltonian \eqref{eq:raman}. For the hyperfine levels $\ket{a}$ and $\ket{b}$ chosen above, the Raman transition in Hamiltonian \eqref{eq:raman} can be achieved by applying two $\hat\sigma_{+}$ polarized Raman fields with different frequencies. As the 2D lattice is on the $x$-$y$ plane, this can be naturally done by applying the light in $\hat z$ direction. Suppose the spatial profiles of the Raman fields are given by $\Omega_{\{a,b\}}(x,y)e^{ikz}$. The Raman transition rate at site $\mb r_{j,k}$ is then given by $\omega_{j,k}e^{i\phi_{j,k}}\propto\int\Omega_{a}^{*}(x,y)\Omega_{b}(x,y)W_{a}(x-jr_{0},y-kr_{0})W_{b}(x-jr_{0},y-kr_{0})dxdy$, where $W_{a}(x,y)$ and $W_{b}(x,y)$ are the Wannier functions for $\ket a$ and $\ket b$, respectively, and the $z$ dependency is assumed to be decoupled. Suppose $\Omega_{a}^{*}(x,y)\Omega_{b}(x,y)\simeq|\Omega_{0}|^{2}e^{i\phi(x,y)}$. We require $\phi(x,y)$ to be slowly varying compared to $W_{\{a,b\}}(x,y)$ so that $\omega_{j,k}$ to be the same for every site. $\phi_{j,k}$ is then approximately given by $\phi(jr_{0},kr_{0})$. Note that any nonlinear term in $\phi(x,y)$ can lead to non-vanishing magnetic flux $(\theta_{j,k}-\theta_{j,k+1})\not=0$. For example, computer generated holography would be relevant for adjusting $\phi(x,y)$ arbitrarily as far as its resolution limit permits \cite{jahns08}. 

A truly arbitrary magnetic field would be possible when $\theta_{j,k}$ (or $\phi_{j,k}$) can be controlled site by site. At first sight, this seems to require diffraction-limited imaging. For example, this is the case for a uniform magnetic field $\phi_{j,k}=2\pi\alpha jk$ (mod $2\pi$) in the sense that for lattice sites with $j=1/2\alpha$, the phase $\phi_{j,k}=(-1)^{k}\pi$ should be flipped at every adjacent site. As will be shown below, however, this seemingly necessary requirement of diffraction-limited imaging is in fact not essential in our scheme, although it should also be noted that subdiffraction imaging technologies are already available \cite{liu07}. For convenience, let us represent the sites $(j,k)$ with a single index $\xi$ or $\lambda$. We apply at each site $\xi$ a moderately well focused Raman field for $\ket{b}$ with spatial profile $\omega_{\xi}'e^{i\phi_\xi'}{\mathcal{A}(\mb{r}-\mb{r}_{\xi})}$, where $\mathcal{A}(\mb{r})$ is a common normalized mode function, e.g., a Gaussian function, while the other Raman field for $\ket{a}$ is applied globally and uniformly. The Raman transition rate at site $\lambda$ is then given by the summation of all the contributions $\omega e^{i\phi_{\lambda}}\propto\sum_{\xi}T_{\lambda\xi}\omega_{\xi}'e^{i\phi_{\xi}'}$, where $T_{\lambda\xi}=\int{\mathcal{A}(\mb{r}-\mb{r}_{\xi})}W_{a}(\mb r-\mb r_{\lambda})W_{b}(\mb r-\mb r_{\lambda})dxdy$. For well focused beams, $T_{\lambda\xi}$ would be non-vanishing only for small $|\mb{r}_{\lambda}-\mb{r}_{\xi}|$. The problem now reduces to determining a proper set of $\omega_{\xi}'$ and $\phi_{\xi}'$ to obtain the desired field profile corresponding to the given set of $\phi_{\lambda}$. It is easily seen that this can be done by calculating the inverse matrix of $T$ for any vector $\omega e^{i\phi_{\lambda}}$, i.e., any $\mb{A}(\mb{r})$. A similar idea has been used in a different context in Ref. \cite{friesen08}. 

In summary, we have shown that in a state-dependent optical lattice, a light field itself can play the role of a gauge field. This is an intriguing situation where the phase gradient of light solely produces a torque to atoms. While this has been known in a different context \cite{babiker94,*roichman08}, it will be an interesting future work to clarify their analogies.

We acknowledge the UK EPSRC for financial support.

\end{document}